\documentclass[structabstract]{aa}

\usepackage{txfonts}
\usepackage{natbib}
\usepackage{graphicx}

\newcommand{\ms}{m s$^{-1}$}

\begin{document}
\bibliographystyle{aa}

\title{GJ 1214b revised}

\subtitle{Improved trigonometric parallax, stellar parameters, 
orbital solution, and bulk properties for the super-Earth GJ 1214b}

\author{Guillem Anglada-Escud\'e\inst{1,2}
  \and B\'arbara Rojas-Ayala\inst{3}
  \and Alan P. Boss\inst{2}
  \and Alycia J. Weinberger\inst{2}
  \and James P. Lloyd \inst{4}
  }

\offprints{G. Anglada-Escud\'e, \email{guillem.anglada@gmail.com}}

\institute{
       Universit\"{a}t G\"{o}ttingen,
       Institut f\"ur Astrophysik,
       Friedrich-Hund-Platz 1,
       37077 G\"{o}ttingen, Germany
  \and Carnegie Institution of Washington,
       Department of Terrestrial
       Magnetism, 5241 Broad Branch Rd. NW, Washington D.C.,
       20015, USA
  \and Department of Astrophysics,
       American Museum of Natural History,
       Central Park West and 79th Street,
       New York, NY 10024
  \and Department of Astronomy, Cornell University, 122 Sciences Drive, Ithaca, NY 14853,
       USA
}

%\shortauthors{Anglada-Escude, Rojas-Ayala et al.}
\date{submitted March 2012}

% Context, Aims, methods, results, conclusions
\begin{abstract}
{GJ 1214 is orbited by a transiting super-Earth-mass planet.
It is a primary target for ongoing efforts to understand
the emerging population of super-Earth-mass planets
around M dwarfs, some of which are detected within the
liquid water (habitable) zone of their host stars.
} {
We present new precision astrometric measurements, a re-analysis of
HARPS radial velocity measurements, and new medium-resolution 
infrared spectroscopy of GJ 1214. We
combine these measurements with recent transit follow-up
observations and new catalog photometry to provide a
comprehensive update of the star--planet properties.
} {
The distance is obtained with 0.6\% relative uncertainty 
using CAPScam astrometry. The new value increases the 
nominal distance to the star by $\sim$10\% and is 
significantly more precise than previous 
measurements. Improved radial
velocity measurements have been obtained analyzing
public HARPS spectra of GJ 1214 using the HARPS-TERRA
software and are 25\% more precise than the original
ones. The Doppler measurements combined with recently
published transit observations significantly refine the
constraints on the orbital solution, especially on the
planet's eccentricity. The analysis of the infrared
spectrum and photometry confirm that the star
is enriched in metals compared to the Sun.
} {
Using all this new fundamental information,
combined with empirical mass--luminosity
relations for low mass stars, we derive updated values
for the bulk properties of the star--planet system.
We also use infrared absolute fluxes to estimate
the stellar radius and to re-derive the star--planet
properties. Both approaches provide very consistent
values for the system. Our analysis shows
that the updated expected value for the planet
mean density is 1.6$\pm$0.6 g cm$^{-3}$, and that a
density comparable to the Earth ($\sim$ 5.5 g cm$^{-3}$)
is now ruled out with very high confidence.
} {
This study illustrates how the fundamental
properties of M dwarfs are of paramount importance in
the proper characterization of the low mass planetary
candidates orbiting them. Given that the distance is now
known to better than 1\%, interferometric measurements
of the stellar radius, additional high precision Doppler
observations, and/or or detection of the secondary transit
(occultation), are necessary to further improve the constraints 
on the GJ 1214 star--planet properties.
}

\end{abstract}

\maketitle

%\email{guillem.anglada@gmail.com}

\section{Introduction}

In the last few years, it has become clear that a large
fraction of low mass stars \citep[$>30\%$ ][]{bonfils:2011} are 
orbited by super-Earth-mass planets on relatively short period orbits,
including a handful detected in the
liquid water (habitable) zones \citep[e.g., GJ 581d and GJ
667Cc: ][respectively]{mayor:2009, anglada:2012b}. These planets have
small but non-negligible probabilities of transiting in
front of their host stars. Given the favorable
star--planet size ratio, we already have the technical
means to begin spectroscopic characterization of their atmospheres. 
GJ 1214b was the first super-Earth
found to transit in front of an M dwarf
\citep{charbonneau:2009} and, as a consequence, it has
received significant observational and theoretical
attention in the past three years. While the orbit is too
close to the star to support hospitable oceans of liquid
water, GJ 1214b offers the first opportunity to attempt
atmospheric characterization of a super-Earth
\citep{bean:2010, croll:2011, berta:2012}, but
apparently contradictory results have
emerged from these studies. Different gases in the
atmosphere of a planet absorb light more efficiently at
certain wavelength ranges. As a result, one should be able to
measure a different effective transit depth as a function
of wavelength. For example, \citet{croll:2011} reported
excess absorption in the K band, which would be
compatible with an extended and mostly transparent
atmosphere with a strong absorber in the near infrared.
On the other hand, \citet{bean:2010} could not detect
significant features in the optical transit depths. This
result was confirmed by the same group in a more extended
wavelength range \citep{bean:2011, desert:2011} and has also been
confirmed using HST spectrophotometric observations in
the near infrared \citep{berta:2012}. Such a flat transmission
spectrum favors a very opaque atmosphere with high
concentrations of water vapor as the main source of
opacity, indicating that the planet could be mainly
composed of water.

All these interpretations rely on atmospheric
models that are strongly dependent on the planet's bulk
properties, especially its mass, radius, mean density,
surface gravity, and stellar irradiation. Prior to the
detection of the planet candidate, \object{GJ 1214} 
was a largely
ignored M dwarf. As a result, some of its fundamental
properties had significant uncertainties (e.g., its
distance). Uncertainties in the fundamental
parameters of GJ 1214 propagate strongly into the planet's bulk
properties, adding an extra element of uncertainty in
discussions about the possible nature of its atmosphere.

Probably the most significant measurement we provide here is
the new measurement of its trigonometric
parallax at 0.6\% precision \citep[previous parallax measurement had a
$\sim$ 10\% uncertainty, ][]{vanaltena:2001}. As a consequence, 
the luminosity
and mass of GJ 1214 have experienced significant updates as
well. The orbital solution for GJ 1214 can also be updated using
recently published transit observations and refined Doppler measurements
obtained with our newly developed software 
\citep[HARPS-TERRA, ][]{anglada:2012a}. In addition, the WISE 
catalog \citep{wise:2011} has also been recently released,
adding four more absolute flux measurements of GJ 1214, thus enabling a
comprehensive spectral energy distribution adjustment and a more secure
determination of T$_{\rm eff}$. In Section \ref{sec:observations}, 
we present the new CAPSCam astrometric measurements
and our re-analysis of public HARPS Doppler
measurements. A new Keplerian solution for GJ 1214b is 
presented in Section \ref{sec:rvorbit}. In Section \ref{sec:star}, we give an
overview of the stellar properties in the light of the
new distance measurement, its near infrared spectrum, and updated
absolute magnitudes. Finally, Section
\ref{sec:starplanet} combines all of the transit
observables with the new orbital solution and provides
the posterior probability distributions for the updated
star--planet parameters. Our conclusions are summarized
in Section \ref{sec:conclusions}.

\section{Observations and Data Reduction}
\label{sec:observations}

\subsection{Astrometry and trigonometric parallax}
\label{sec:astrometry}

The trigonometric parallax of GJ 1214 has
been obtained using the Carnegie Astrometric Planet
Search Camera \citep[CAPSCam,][]{boss:2009} installed in 
on the 2.5m duPont
telescope of the Las Campanas Observatory (Chile). 
The observations span 26 months (June 2009 to Sep 2012)
and 10 epochs have been obtained at an average precision
of 1.0 milliarcsec (mas) per epoch.
Each epoch consists of 20 or more exposures of
45 seconds each. GJ 1214 is significantly brighter than
the average background sources and would saturate the detector
in less than 10 seconds. However CAPSCam can read out
a small part of the array much faster than the full
field \citep{boss:2009}. For GJ 1214, a window of
64$\times$64 pixels is read out every 5 seconds, and all
subrasters are added to the final full field image. The
field of view of CAPSCam is 6.6 arcminutes wide and is
typically rich in background stars, so a very robust reference
frame with more than 30 objects can be used to correct
for field distortions. Centroid extraction, source
crossmatching, field distortion correction, and
astrometric solutions for all the stars in the field have
been obtained using the ATPa astrometric software
developed within the CAPS project (available upon
request). The methods and algorithms applied are outlined
in \citet{boss:2009} and \citet{anglada:2011}.

A measurement of the parallax and proper motion requires
fitting 5 astrometric parameters simultaneously: initial
offset in $R.A.$ and $Dec.$, proper motion in $R.A.$ and $Dec.$,
and the parallax itself. The formal
uncertainties from a classic least-squares analysis
(e.g., from the diagonal of the covariance matrix) are
usually overoptimistic and do not properly account for
correlations between parameters. To overcome these issues, we
estimate the uncertainties in the astrometric parameters
using a Monte Carlo approach. This is done by generating many
realistic sets of measurements and measuring the standard
deviation of the resulting derived parameters for the entire
Monte Carlo-generated sample.

\begin{figure}[tb]
\centering
\includegraphics[width=3.0in,clip]{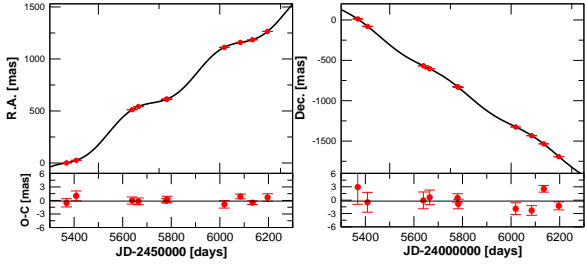}

\caption{Astrometric motion of GJ 1214 as a function of
time. Top panels represent the motion in R.A.(left)
and Dec. (right). Bottom panels show the residuals to
the astrometric fit. The RMS of the residuals is
below 1 mas.} \label{fig:astrometry}

\end{figure}

To do this properly, a first realistic estimate of the epoch-to-epoch accuracy
is needed. The outputs of the astrometric processing are the astrometric
parameters of the target star and of all the other objects in the field. By
combining the residuals of all the reference stars from all the epochs, we can
compute the expected uncertainty per epoch. Doing this, we obtain an
epoch-to-epoch accuracy of $\sim$ 1.3 mas on both $R.A.$ and $Dec.$. Since the
reference stars are fainter than the target, this is a conservative estimate of
the real precision for the target. The star itself, which is not included in the
reference frame, shows a standard deviation in the residuals of 1.0 mas/epoch,
which is also consistent with the expected CAPSCam performance, assuming 20
minutes of on--sky observations per epoch \citep[see][]{boss:2009}. Then, we
simulate 10$^5$ synthetic sets of astrometric observations (same format as Table
\ref{tab:astrodata}) using the nominal parallax and proper motion at the same
epochs of observation. Random Gaussian noise with a single epoch uncertainty of
1.3 mas is injected into each synthetic data set and the 5
astrometric parameters are derived. The standard deviation of each parameter
over the 10$^5$ solutions is the corresponding uncertainty. The
obtained uncertainty for the differential parallax is 0.44 mas and corresponds
to a relative precision of $\sim0.6\%$. We note that this Monte Carlo method
implicitly accounts for parameter correlation \citep[see discussion
in][]{faherty:2012}. The differential astrometric measurements used to measure
the differential parallax and proper motion of GJ 1214 are given in Table
\ref{tab:astrodata}. The best fit solution over-plotted with the astrometric
epochs is shown in Figure \ref{fig:astrometry}.

\begin{table}
\caption{Local astrometric measurements of GJ 1214}\label{tab:astrodata}
\begin{tabular}{lrrrr}
JD      & R.A.  & $\sigma_{\rm R.A.}$ & Dec.  & $\sigma_{\rm Dec.}$ \\
(days)  & (mas) & (mas)             & (mas) & (mas)              \\
\hline\hline
 2455368.781418 &      0.45  &    0.91   &     12.30 &  3.86   \\ 
 2455408.647011 &     25.62  &    1.09   &    -78.97 &  2.22   \\
 2455638.911225 &    511.34  &    0.79   &   -567.61 &  1.88   \\
 2455664.846719 &    542.31  &    0.74   &   -604.85 &  1.64   \\
 2455779.634773 &    611.31  &    0.45   &   -826.25 &  0.93   \\
 2455782.624045 &    614.41  &    0.67   &   -834.84 &  1.07   \\
 2456018.887788 &   1111.53  &    0.81   &  -1325.39 &  1.32   \\
 2456084.702758 &   1159.22  &    0.52   &  -1433.52 &  1.08   \\
 2456134.624810 &   1184.17  &    0.39   &  -1534.43 &  0.74   \\
 2456196.509776 &   1264.98  &    0.82   &  -1692.83 &  0.88   \\
\hline
\end{tabular}
\end{table}

The measured parallax is relative to the
reference stars. Since they are not at infinite
distances, these reference stars also have parallactic motion. As a
result, the average parallax of the reference frame
cannot be derived from the astrometric observations
alone. We obtain this average parallax (also called the
parallax zero-point correction) using catalog photometry
for the reference stars using the procedure described in
\citet{anglada:2011} with 19 reference stars with good
CAPSCam astrometry (RMS of the residuals below 2.0
mas/epoch) and reliable catalog photometry
\citep[$B$$<$18 NOMAD and $JHK$$_s$ 2MASS photometry;][]{nomad,twomass}.
For this set of observations, we determine
a zero--point correction of -0.40 $\pm$ 0.3 mas. This value,
when added to the relative parallax, produces
the final absolute parallax measurement in Table
\ref{tab:astroresults}. The inverse of the parallax in arcseconds
is the distance in parsecs (pc) and amounts to
14.55 $\pm$ 0.13 pc. The updated $BVRIJHK_s$, $W1,W2, W3$
and $W4$ absolute magnitudes for GJ 1214 are presented and
discussed in Section \ref{sec:star}.

\begin{table}
\caption{Basic astrometric information and
results from the analysis of the astrometry.}
\label{tab:astroresults}
\centering                                      % used for centering table
\begin{tabular}{lrr}          % centered columns (4 columns)
Parameter  & Value \\
\hline\hline                        % inserts double horizontal lines
R.A.		             &  17 15 18.94\tablefootmark{1}  \\
Dec.		             & +04 57 49.7 \tablefootmark{1}  \\
Catalog $\mu^*_{\rm R.A.}$ [mas yr$^{-1}$]
                             & 585 \tablefootmark{2} \\
Catalog $\mu_{\rm Dec.}$ [mas yr$^{-1}$] 
                             & -752 \tablefootmark{2}  \\
\hline\hline
Relative  $\mu^*_{\rm R.A.}$ & 581.88   $\pm$   0.5  \\
Relative  $\mu_{\rm Dec.}$   & -734.6   $\pm$   0.8  \\
Relative parallax [mas]      &  69.11   $\pm$   0.4  \\
Zero-point correction [mas]  &  -0.4    $\pm$   0.3  \\
Absolute parallax [mas]      &  68.71   $\pm$   0.6  \\
%Distance [pc]                &  14.47   $\pm$   0.18 \\
Distance [pc]                &  14.55   $\pm$   0.13 \\
\hline
\end{tabular}
\tablefoot{
\tablefoottext{1}{2MASS catalog \citep{twomass}}
\tablefoottext{2}{LSPM-NORTH catalog \citet{lepine:2005}}
}
\end{table}

We note that the measured proper motions are also
differential and contain an unknown offset due to the
unknown average motion of the background stars (e.g., all
galactic plane stars move roughly in the same direction).
Even though catalog values are less precise than the
ones we obtain, catalog proper motions are usually
corrected for proper motion zero--point ambiguities.
Differential measurements as well as
the suggested values for the proper motion of GJ 1214
\citep[LSPM-NORTH catalog, ][]{lepine:2005} are also
provided in Table \ref{tab:astroresults}.

\subsection{Radial velocity measurements} \label{sec:rvdata}

The ESO public archive\footnote{
\texttt{http://archive.eso.org/wdb/wdb/eso/repro/form}}
contains 21 reduced HARPS spectra of GJ 1214 obtained by
\citet{charbonneau:2009}. We reanalyzed these public
HARPS spectra using our software called HARPS-TERRA
\citep[HARPS Template Enhanced
Radial velocity Re-analysis Application,][]{anglada:2012a}. 
HARPS-TERRA derives
differential RV measurement by constructing a high
signal-to-noise ratio template from the observations and
matching it to each spectrum using a least-squares
approach. GJ 1214 is faint at optical wavelengths
(V$\sim$14). As a consequence, very long exposures (2400
sec) were required to obtain any signal at all. Still,
the typical S/N at 6100 \AA\ is only $\sim$10 so 
careful construction of the template is a key element in
order to achieve the maximum precision. The standard setup
of HARPS-TERRA for M dwarfs uses all the echelle
apertures redder than the 22nd one (4400\AA $< \lambda
<$ 6800\AA) and adjusts a cubic polynomial to correct
for the variability of the blaze function across each
echelle order. Detailed algorithms and
performance of HARPS-TERRA on a representative sample of
stars is given in \citet{anglada:2012a}. It is
worth noticing that error bars
listed in \ref{tab:rvdata} are smaller than those in
\citet{charbonneau:2009}, suggesting a more optimal
usage of the Doppler information in the spectra. However,
reduced error bars do not guarantee a more 
accurate orbital fit when the RV variability
is dominated by systematic noise (either instrumental 
or stellar). To account for this, we included an 
unknown noise term in our Monte Carlo Markov Chain 
runs which had a substantial effect on the derived 
distributions, especially on the orbital eccentricity 
(see discussion in Section \ref{sec:rvorbit}). 

\begin{table}

\caption{HARPS-TERRA radial velocity measurements.}
\label{tab:rvdata}

\centering                                      % used for centering table
\begin{tabular}{lcccc}
JD & RV  & $\sigma_{RV}$ \\
(days) & (\ms) & (\ms) \\
\hline\hline\\
2455036.57372   &  -10.82 &  1.80\\
2455036.65153   &   -4.77 &  2.08\\
2455037.58578   &    6.86 &  3.63\\
2455037.65309   &   -3.37 &  1.72\\
2455038.53985   &    5.67 &  1.41\\
2455038.63702   &    5.61 & 1.78\\
2455039.55202   &  -14.88 & 1.69\\
2455039.63876   &  -13.53 &  1.61\\
2455040.56221   &   14.15 &  1.84\\
2455040.63961   &    5.90 &  2.49\\
2455041.57417   &    5.68 &  1.70\\
2455042.52391   &   -2.11 &  3.91\\
2455042.54566   &   -5.57 &  1.37\\
2455042.63521   &   -4.15 &  1.42\\
2455045.55962   &    3.43 &  2.92\\
2455045.64403   &    1.61 &  2.52\\
2455046.55684   &   10.59 &  1.97\\
2455046.63141   &    8.28 &  1.94\\
2455047.55042   &  -13.26 &  2.19\\
2455048.54997   &    1.64 &  1.93\\
2455048.61096   &    3.03 &  1.63\\
\end{tabular}
\end{table}

\subsection{Orbital solution}\label{sec:rvorbit}

As discussed in \citet{carter:2011}, uncertainties in the
orbital parameters (especially the eccentricity) are the main
limitation in the derivation of precise star--planet
parameters. Therefore, we are interested not only in the
favoured orbital solution, but also in obtaining a realistic
numerical representation of the posterior density function of
the parameters to propagate them into the estimates of the
star--planet properties. These samples are generated using a 
Bayesian Monte Carlo Markov chain (MCMC) method.
We use custom-made MCMC software to combine RV with transit
observations and to obtain such distributions. The software is
based on the MCMC tools used in \citet{anglada:2011} adapted
to the Doppler plus transit problem, and uses a Gibbs sampler. Jump
scales of the Gibbs sampler are initialized using the 
diagonal elements of the covariance matrix at the maximum
likelihood solution, and they are
adjusted in the first 10$^5$ steps (the ``burn-in'' period) to accept
between 10\% and 30\% of the jump proposals for each parameter.
As discussed below, we assume flat priors for all the sampling
parameters. As a consequence, the probability of accepting a
new state is just the ratio of likelihoods at the current and
proposed jump position. An outline of the general MCMC method
applied to the Keplerian problem can be found in
\citet{ford:2006}.

The Keplerian model for the radial velocities is the same as
in \citet{anglada:2012c} and the transit observations are
predicted using the recipes given in the NASA Exoplanet
Archive webpage 
\footnote{http://exoplanetarchive.ipac.caltech.edu/applications/TransitSearch/
guide/algorithms.html}. 
Transit observations strongly constrain the orbital period, but also put
strong restrictions on the sum of the initial mean anomaly
$M_0$, the argument of the node $\omega$ (also called, mean
longitude $\lambda_0=M_0+\omega$), and the product
$e\cos\omega$ (eccentricity and argument of the node).
Accordingly, we use $\lambda_0$ as a free parameter. Since the
transit instants explicitly depend on $e\cos \omega$,  $e\sin
\omega$ and $e \cos \omega$ are sometimes used as the MCMC
sampling parameters. We also tested this approach,
and, while it has some desirable properties (e.g., symmetry
around zero eccentricity)-- this parameterization imposes an
\textit{implicit prior} \footnote{The implicit prior comes
from the fact that $x=e\sin\omega$ and $y=e\cos\omega$ are
cartesian coordinates derived from the polar radial coordinate
$e$ and the angle $\omega$. Because MCMCs are based on the
integration of a probability distribution, to preserve the
flat prior choices for $e$ and $\omega$, one would need to
include the Jacobian of the transformation ($1/e$) as a prior
when using $x$ and $y$ as the Markov chain sampling parameters.
The use of such a prior is undesirable due to numerical
stability issues (divergence for $e=0$).} that severely biases the
estimates of $e$ to artificially larger values \citep[e.g.,
see ][for similar discussions]{ford:2006, barros:2011}.
Therefore, at the sampling level, we choose those free
parameters that, from our point of view, should have flat
prior distributions. These are: orbital frequency $1/P$, RV
semi-amplitude $K$, mean longitude $\lambda_0$, argument of
the node $\omega$, and eccentricity $e$. Despite $\omega$ being
poorly defined and uncertain at low eccentricities, the
convergence of the MCMC was substantially improved thanks to
the use of $\lambda_0$ instead of $M_0$, so we find this
parametrization convenient and sufficient for our purposes. 
All the approximate parameter values are already known from
previous studies and, therefore, we initialize the MCMCs
within 3 standard deviations of the orbital solution proposed
by \citet{charbonneau:2009}.

Real uncertainties in radial velocity measurements are
difficult to estimate properly, especially when the star is
active. To account for that, we model the uncertainty of the
$i$-th measurement as $\sigma_i^2=\epsilon_i^2+s^2$, where
$\epsilon_i$ are the formal uncertainties derived from
HARPS-TERRA (third column in Table \ref{tab:rvdata}), and $s$ is a
\textit{jitter parameter} that accounts for the extra systematic
noise. In this context, the jitter parameter $s$ is treated
as any other free parameter.

We assign a constant uncertainty to each transit instant of
400 sec. This uncertainty is 10 times larger than the typical
reported formal errors but it is more consistent with the
different instants of transit measurements obtained
simultaneously by different groups (see Table
\ref{tab:transittimes}). While the MCMC convergence properties
are greatly improved, 400 sec is a small fraction of the
time-baseline covered by the transits and, as a consequence,
the period is still strongly constrained. The transit time
measurements have been extracted from \citet{sada:2010},
\citet{carter:2011}, and \citet{kundurthy:2011}.

\begin{figure}[tb]
\centering
\includegraphics[width=3.0in,clip]{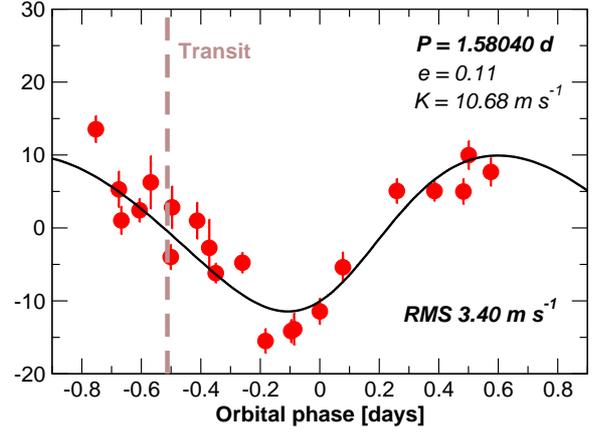}

\caption{HARPS-TERRA radial velocity measurements
folded to the orbital period. The instant of transit
is depicted as a dashed vertical line. } \label{fig:rvfit}

\end{figure}

\begin{figure}[tb]
\centering
\includegraphics[width=3.0in,clip]{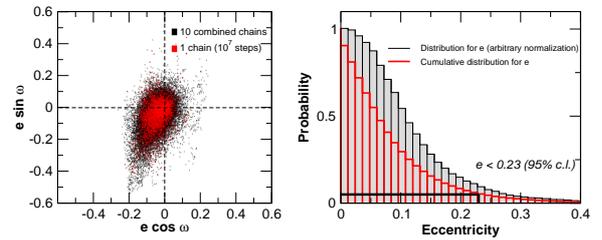}

\caption{\textbf{Left.} Distribution of MCMC states for the
derived quantities $e \cos \omega$ and $e \sin \omega$ as
obtained from 1 MC chain of $10^7$ steps (red) and
the combination of 10 chains of 10$^7$ steps each (black).
Only $1\%$ of the steps are included in this plot to improve
visualization. \textbf{Right.} Marginalized probability
distribution of the eccentricity is shown in black (arbitrary
normalization). The corresponding cumulative distribution (red
histogram) shows that eccentricities higher than 0.23 are
ruled out at a 95\% confidence level.}

\label{fig:edist}

\end{figure}

In table \ref{tab:rvsolution}, we provide the expected values
for the most relevant combination of parameters for physical
and transit prediction computations as derived from the 
combination of 10 MCMC runs with $10^7$ steps each. Figure
\ref{fig:rvfit} is an illustration of the maximum \textit{a posteriori}
probability (MAP) solution and is given only for illustration
purposes. Since the amplitude of the signal is small
compared to the noise, the eccentricity is still poorly
constrained. The posterior distribution of the eccentricity has
a maximum very close to $0$ and monotonically decreases with
$e$ (see Figure \ref{fig:edist}) implying that close to
circular orbits are favored. Because of their significance
for the transit observables (e.g., see Section 
\ref{sec:starplanet}), an illustration of the 
MCMC samples of the derived parameters
$e \cos \omega$ and $e \sin \omega$ is also provided in
figure \ref{fig:edist}. Previous estimates of the
eccentricity of GJ 1214b \citet[e.g., ]{charbonneau:2009,
carter:2011} seemed to favor eccentricities close to 0.1, but
still compatible with 0. In initial tests, we recovered this
same slightly higher eccentricity value if the noise parameter
$s$ was fixed to $0$. This indicates that previous estimates
of the orbital elements were biased due to a non-consistent 
treatment of the RV uncertainties. Even though stellar 
jitter dominates the error budget ($s \sim$ 2.9 \ms), 
the RMS of the MAP solution using the
new HARPS-TERRA measurements is lower (3.4 \ms) than the one
reported by \citet[][4.4 \ms]{charbonneau:2009}. Given the
strong effect of the jitter parameter $s$ on the result, and
until more RV measurements become public, we strongly
recommend the use of the updated solution in Table
\ref{tab:rvsolution} for any future work (e.g., in searching for
secondary transits). Applying the same 95\% confidence level
used by \citet{charbonneau:2009}, we obtain an upper limit to
the eccentricity of 0.23. 

\begin{table}

\caption{Expected values of the most useful parameter
combinations obtained using RV and transit observations. Numbers 
in parentheses represent the standard deviation of the 
distributions (last two significant digits of the expected 
values).}
\centering
\label{tab:rvsolution}
\begin{tabular}{lrr}
Parameter &
\\
\hline\hline\\
P [days]                       &  1.580400(14) \\
K [\ms]                        &  10.9(1.6)    \\
$\lambda_0$ [deg]              &  210.9(6.2)\tablefootmark{a} \\
e$\cos \omega$            &  -0.033(55)   \\
e$\sin \omega$            &  -0.044(90)   \\
Jitter s [\ms]                 &  3.6(1.1)     \\
$\gamma$                       &  0.35(1.2)    \\
\\
Other parameter combinations\\
\hline
e                              &  $<$0.23\tablefootmark{b}       \\
$M \sin i$ [$M_{jup}$]         &   0.0195(28)\tablefootmark{c} \\
$M \sin i$ [$M_{\oplus}$]      &   6.20(91)\tablefootmark{c} \\
a [AU]                         &   0.0148\tablefootmark{c}     \\
\hline
\\
\hline\\
N$_{transits}$                 &  30        \\
N$_{RV}$                       &  21        \\
RMS$_{RV}$ [\ms]               &  3.5       \\
\hline                                                            \\
\hline                                             %inserts single line
\end{tabular}
\tablefoot{
\tablefoottext{a}{
Julian date of the reference epoch on which $\lambda_0$ 
is computed is the first epoch of the RV data at 
$T_0=2455036.57372$ days. For a circular orbit $\lambda_0$ 
is equal to the mean anomaly $M_0$ at the reference epoch.
}
\tablefoottext{b}{Distribution of the eccentricity peaks 
close to $0$ (see Figure \ref{fig:edist}).
}
\tablefoottext{c}{Assumes M$_*$=0.176 M$_\odot$. The
uncertainty in the mass of GJ 1214 is considered
in the final proposed parameters for the star--planet
system given in Section \ref{sec:starplanet}.
}
}
\end{table}

%\begin{figure}[tb]
%\centering
%\includegraphics[width=3.0in,clip]{all_ecosw.eps}

%\caption{\textbf{Left.} Distribution of MCMC states for $e \cos \omega$ 
%and $e \sin \omega$ obtained from 1 MC chain of $10^7$ steps 
%(red) and the combination of 10 chains of 10$^7$ steps each (black). 
%\textbf{Right.} Marginalized distribution of $e \cos \omega$, which
%is the mostly constrained component of the eccentricity 
%imposed by the transit times.} 
%\label{fig:rvfit}

%\end{figure}

Another interesting parameter is the minimum mass of GJ 1214b.
Even though the new RV amplitude $K$ is smaller ($\sim 10.9$
\ms) than the previously reported one ($\sim$12.5 \ms), the
updated minimum mass does not change significantly compared to
\citet{charbonneau:2009} due to the similar relative increase
in the updated stellar mass derived from the new distance
determination (see Section \ref{sec:starplanet}). The origin
of the slightly smaller amplitude is likely caused by stellar
activity. By analyzing Doppler measurements on known active M
dwarfs \citep[e.g., AD Leo, ][]{reiners:2012b}, we found that the
classic HARPS-CCF approach always produces a larger Doppler
amplitude than the one derived using HARPS-TERRA (between 1.1
to 1.5 times larger) if a candidate signal is caused by
activity. The likely explanation for this is the different
sensitivities of the two methods to the changes in the stellar
line-profiles. While one could use this effect to obtain a
further diagnostic to check the reality of low-mass candidate
planets (under investigation), in this case it means that the
most likely source of systematic noise comes from the star
rather than from the instrument. Doppler follow-up of the star
is needed to better understand the origin of the extra
noise and to further constrain the orbital solution.

\section{Updated properties for GJ 1214} \label{sec:star}

\begin{table}
\caption{Updated absolute photometry for GJ 1214.
Johnsons-Cousins BVRI photometry is from
\citep{dawson:1992}. JHKs photometry is from the 2MASS
catalog \citep{twomass}. The W1, W2, W3 and W4 are the
mid-infrared bands from the WISE
Preliminary Data Release \citep{wise:2011}
(central wavelengths are 3.4 $\mu$, 4.6$\mu$m,
12$\mu$m and 22$\mu$m). Properties derived from
the fits of the absolute photometry to the
BT-Settl-2010 model grid are also provided.}
\label{tab:magnitudes}
\centering                                      % used for centering table
\begin{tabular}{lrr}
\hline\hline
M$_B$               & 15.532 $\pm$ 0.080\\
M$_V$               & 13.822 $\pm$ 0.041\\
M$_R$               & 12.441 $\pm$ 0.043\\
M$_I$               & 11.608 $\pm$ 0.042\\
\\
M$_J$               &  8.934 $\pm$ 0.041\\
M$_H$               &  8.274 $\pm$ 0.041\\
M$_{Ks}$              &  7.964 $\pm$ 0.038\\
\\
M$_{W1}$	    &  7.781 $\pm$ 0.041\\
M$_{W2}$	    &  7.614 $\pm$ 0.039\\
M$_{W3}$	    &  7.407 $\pm$ 0.041\\
M$_{W4}$            &  7.204 $\pm$ 0.178\\
\hline                                             %inserts single line
T$_{\rm eff}$ [K]	& 3252 $\pm$ 20\tablefootmark{1}    \\
$L_*$ [10$^{-3}$ $L_\odot$] & 4.05 $\pm$ 0.19               \\
R$_*$ [$R_\odot$]           &  0.201  $\pm$ 0.010           \\
$[$Fe/H]$_{phot} $ 	& $+$0.13, $+$0.05\tablefootmark{2}    \\
$[$Fe/H]$_{spec} $ 	& $+$0.20\tablefootmark{3}             \\
$[$M/H]$_{spec} $ 	& $+$0.15\tablefootmark{3}             \\
\hline
\end{tabular}
\tablefoot{
\tablefoottext{1}{Uncertainty in T$_{eff}$ only accounts for statistical
errors.}
\tablefoottext{2}{\citet{sl:2010, neves:2012}}
\tablefoottext{3}{\citet{rojas:2012}}
}
\end{table}

\subsection{Metallicity}\label{sec:metallicity}

Two independent techniques, the photometric method by \citet{sl:2010} and the
K-band spectroscopic [Fe/H] index by \citet{rojas:2012}, agree on the
metal-richness of GJ 1214, with [Fe/H] = +0.28 dex and [Fe/H] = +0.20 dex,
respectively. However, since the [Fe/H] photometric calibrations depend on the
distance of the star (to obtain M$_{K_s}$), the photometric value of GJ 1214
needs to be re-calculated using its updated distance. The updated absolute
magnitudes are given in Table \ref{tab:magnitudes}. The updated photometric
[Fe/H] values are [Fe/H] = +0.13 dex and [Fe/H] = +0.05 dex, using the
\citet{sl:2010} calibration and the recent calibration by \citet{neves:2012},
respectively. Considering the dispersions associated with the calibrations
($\sigma$~ 0.1-0.15 dex), the estimates are consistent with each other,
corroborating that GJ 1214 is a solar or super-solar [Fe/H] star.

\begin{figure}[tb]
\centering
\includegraphics[width=3.2in,clip]{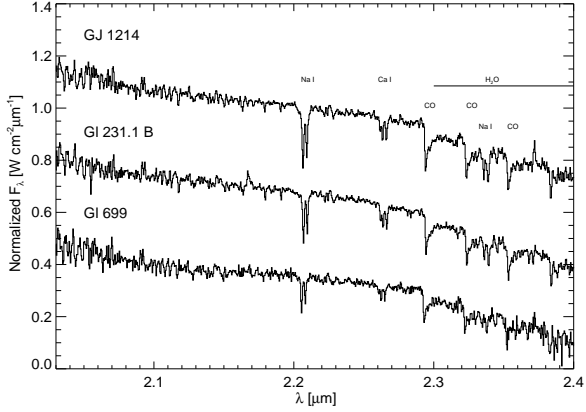}

\caption{K-band spectra of GJ 1214 (top), Gl 231.1B (middle),
and Gl 699 (bottom). All the stars have the same K-band
spectral type (same overall shape of their spectra), but
different metallicities, as indicated by the strengths of
their absorption features. GJ 1214's metallicity should be at
least equal to or higher than solar.} \label{fig:kbandspectra}

\end{figure}

A comparative approach can also be performed to confirm the
metal-richness of GJ 1214. Figure \ref{fig:kbandspectra} shows the
K-band spectra of Gl 699 (Barnard's star), Gl 231.1B, and
GJ1214\footnote{These spectra are part of the K-band spectral atlas
by \citet{rojas:2012}, and available to the community in the online
version of that article.}. Gl 699 is the second nearest M star to the
solar system and its kinematics, H$_\alpha$ activity, and [Fe/H]
measurements are consistent with Gl 699 being an old disk/halo star
\citep[][]{gizis:1997,rojas:2012}. Gl 231.1B is the low-mass
companion of a nearly solar metallicity G0V star ([Fe/H]=-0.04 dex,
\citet{spocs:2005}). The overall shapes of the K-band spectra of all
three stars are quite similar (same spectral type), and the
morphology of the surface-gravity-sensitive CO bands corresponds to M
dwarf stars (log g $\sim$ 5). However, all the absorption features of
GJ 1214 are stronger than the ones exhibited by metal-poor Gl 699
([Fe/H] = -0.39 dex, \citet{rojas:2012}) and solar metallicity Gl
231.1B. Therefore, also in a relative sense, the strong absorption
features favor a high metal content for GJ 1214.

\subsection{Stellar temperature, luminosity and radius
from models}\label{sec:TLR}

We first provide an overview of previous methods and
determinations of the temperature, luminosity, and
radius of GJ 1214. In the discovery paper of GJ1214b,
\citep{charbonneau:2009} estimated a T$_{\rm eff}$=3026K
and a R$_{\star}$= 0.211R$_{\odot}$ for GJ 1214.
\citet{kundurthy:2011} obtained
T$_{\rm eff}$=2949K and R$_{\star}$= 0.211R$_{\odot}$.
Constraining parameters to stellar evolution isochrones
by \citet{baraffe:1998}, \citet{carter:2011} led to estimates of
T$_{\rm eff}$=3170K and R$_{\star}$= 0.179R$_{\odot}$,
assuming that GJ 1214 is an ``old star''. However, all
these results were based on the distance determination
of $\sim$ 13 pc by \cite{vanaltena:2001}, and reveal
that the effective temperature of GJ 1214 is quite
uncertain, given all the degeneracies in the evolutionary
model used and a low precision distance estimate.

We attempted to get an estimate of the stellar radius using
up-to-date interferometric empirical calibrations. For
example, \citet{kervella:2004} provides an empirical
luminosity--radius calibration that extends to low mass
stars and uses absolute magnitudes as the only inputs. We
tried different combinations of photometric bands obtaining
inconsistent results. All the relations involving optical
bands ($BVRI$) gave values larger than 0.4 R$_\odot$, which
are very unrealistic considering the spectral type of GJ
1214. The relations restricted to infrared colors($J, H$,
and $K$) provided estimates a bit more realistic (between
0.15 to 0.2 R$_\odot$) but still not fully compatible with
each other. We also tried the calibration provided by
\citet{demory:2009} that used K band photometry to minimize
the metallicity effect on optical magnitudes. In
\citet{demory:2009}, new measurements of 6 new M dwarfs
were presented and a new radius--luminosity calibration was
derived. They found a remarkable agreement with the
evolutionary models of \citet{baraffe:1998} if the measured
radii was plotted against the K absolute magnitude. Using
this approach, we obtain a stellar radius of 0.193
R$_\odot$ which, at least, seems to be in the expected
range.

\begin{figure}[tb]
\centering
\includegraphics[width=3.5in,clip]{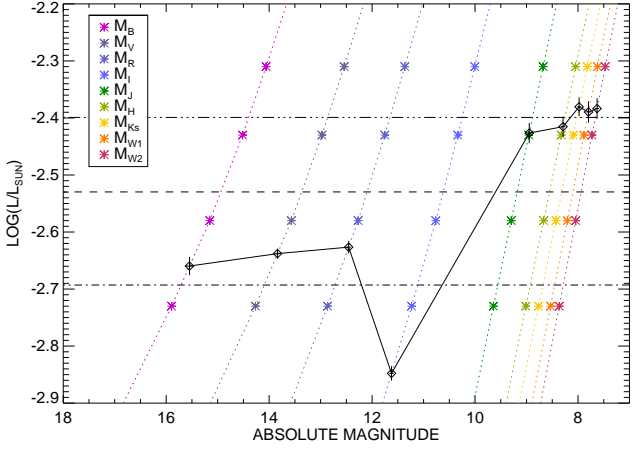}

\caption{Luminosity as function of absolute magnitude for all the
wavebands listed in Table \ref{tab:magnitudes}. Open diamonds
represent the photometry of GJ 1214. The color asterisks represent
the BT-Settl-2010 values, and the color dotted lines the linear
interpolations between them. The dashed line indicates the mean
bolometric luminosity estimated using all the photometry
(log(L/L$_\odot$)= -2.57). The dotted-dashed line indicates the mean
luminosity with only the $BVRI$ photometry (log(L/L$_\odot$)= -2.73),
and the 3 dotted-dashed line the adopted mean luminosity for GJ 1214,
derived using only the $JHK_sW1W1$ absolute magnitudes
(log(L/L$_\odot$)= -2.44).  The lowest bolometric luminosity
corresponds to the I magnitude, where the SED is very steep.}

\label{fig:sed1}
\end{figure}

\begin{figure}[tb]
\centering
\includegraphics[width=3.5in,clip]{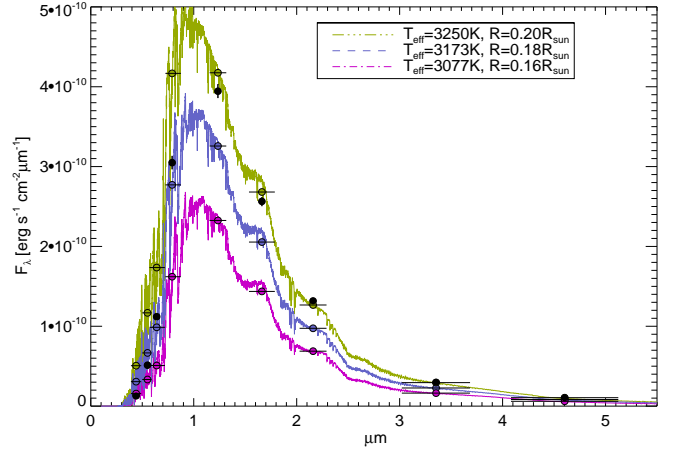}

\caption{Spectral energy distribution fits for GJ 1214 for
the effective temperatures and radii derived from
photometry. The magenta, green, and blue SEDs are the
spectral templates that represent the values obtained with
only BVRI photometry, only $JHK_sW1W2$ photometry, and the
9 wavebands, respectively. All spectral templates have
solar metallicity, log g = 5.0 and distance equal to 14.47
pc. The photometry data of GJ 1214 are plotted as black
dots. Color dots indicate flux levels of each spectral
template integrated over the corresponding filter
bandwidth, depicted by the horizontal black lines. The
spectral template with T$_{eff}$=3250K and R=0.2R$_\odot$
provides the best fit to the infrared photometry of GJ
1214.}

\label{fig:sed2}
\end{figure}

Given that the state-of-the-art empirical relations are not
entirely self consistent, we also obtained a new estimate of the
effective temperature, luminosity, and radius of GJ 1214
using updated $BVRIJHK_sW1W2$ absolute photometry (Table
\ref{tab:magnitudes}) together with the model atmosphere
grid BT-Settl-2010
\citep{allard:2011}\footnote{http://phoenix.ens-lyon.fr/simulator/index.faces}.
The lack of indicators of youth in GJ 1214, such as
H$_\alpha$ emission, along with its space motion, gives an
age estimate of 3--10 Gyr for the star \citep{reid:2005}.
This age estimate is also supported by its long rotation
period \citep{charbonneau:2009}. Therefore, we fixed the
surface gravity of the synthetic spectral models to log g =
5.0 and [M/H]=0.0. Figure \ref{fig:sed1} shows the
bolometric luminosity as function of absolute magnitude for
the BT-Settl-2010 grid. The inferred bolometric
luminosities with the JHKsW1W2 photometry are consistent
with each other, and higher than the luminosities obtained
with the BVRI photometry. Previous synthetic models
\citep[e.g. NextGen;][]{nextgen:1999} showed a lack of flux
in the K band when compared with observed spectra of
low-mass stars. New solar abundances and the inclusion of
dust grain formation seems to have solved most of the
previous discrepancy, allowing the BT-Settl-2010 models to
reproduced fairly well the infrared spectral energy
distribution (SED) of M-dwarfs \citep[][however, the FeH
opacity data is still incomplete for this
region]{allard:2011}. Although models have also
been improved at shorter wavelengths ($BVRI$), they remain
too bright in the ultraviolet and visible part of the M
dwarf spectra, possibly due to missing sources of opacity
in the modeling process \citep{allard:2011}. The
effective temperatures and radii corresponding to the best
SEDs fit to the 9 wavebands $BVRIJHK_sW1W2$, only $BVRI$, and
only $JHK_sW1W2$, obtained from their respective
bolometric luminosities in Figure \ref{fig:sed1}, are shown
in Figure \ref{fig:sed2}. Given that 1) bolometric
luminosities obtained with $JHK_sW1W2$ photometry are
consistent with each other, 2) the models provide a better fit
at longer wavelengths, and 3) the empirical luminosity--radius
provides more self-consistency using nIR colors, the fit with
only $JHK_sW1W2$ provides the most reliable results and is
the one to be used in deriving further properties of the
star--planet system.

The SED fitting using only these nIR bands is further supported
by other results. For example in Section
\ref{sec:starplanet}, we will obtain an independent stellar
radius estimate ($0.211 \pm 0.011$ R$_\odot$) combining the
stellar mass with direct observables from the light curve
and Doppler data. Such an estimate is compatible with the
value we discussed in this section ($\sim$0.2 R$_\odot$). Also,
the estimate of the effective temperature (T$_{\rm eff}\sim
3250 K$) is in excellent agreement with the effective
temperature derived from water absorption in $K$-band by
\citet[][T$_{\rm eff}=$3245 K]{rojas:2012}. Furthermore,
next Section shows this T$_{\rm eff}$ also agrees with the
one derived from SED fits to the evolutionary models
\citep{baraffe:1998}.

\subsection{Stellar mass}\label{sec:empmass}

Although most of the previously reported mass-radius estimates
of GJ 1214 agree, we note that they all assume the
trigonometric distance estimate from \citet{vanaltena:2001}.
This distance determination has an uncertainty of 10\% which
is not always accounted for in the aforementioned predictions.
We also note that the M1V-M4V spectral type range spans a
broad range of stellar masses (from 0.4 to 0.15 M$_\odot$) and
effective temperatures (3500 to 2800 K). This makes any mass
estimate very sensitive to uncertainties in the parallax mesurement. 
Given the updated distance, we can apply the
\citet{delfosse:2000} (hereafter DF00) relations to derive a
new mass for GJ 1214 using the absolute $J$, $H$ and $K$ magnitudes.
The masses derived from each photometric band are M$^{(J)}$ =
0.174 M$_\odot$ ,M$^{(H)}$= 0.177 M$_\odot$ and M$^{(K)}$=
0.177 $_\odot$, which are in very good agreement.

Another way to obtain an estimate of the mass for GJ 1214
is to compare the absolute magnitude with the synthetic
colors from the evolutionary models in
\citet{baraffe:1998}. Unfortunately, \citet{baraffe:1998}
does not contain models for stars with [Fe/H]$>$0.0. As a
result the optical colors (e.g., $B, V, R, I$), cannot be
properly adjusted to a model. The fit including the $BVRIJHK$
magnitudes has T$_{eff}$ = 2987 K ($M$ = 0.12 $M_{\odot}$
and is of very low quality ($\chi^2$ = 45 for 6 degrees of
freedom). The fit to $BVR$ alone give T$_{eff}$ = 2880 K,
$M$ = 0.11 $M_{\odot}$ and a $\chi^2$ = 10.0 for 3 degrees of
freedom. In contrast, the best fit model obtained from
adjusting $J$, $H$, and $K$ has a $\chi^2$ of 1.92 for 3 degrees of
freedom, an effective temperature of 3225 K and a mass of
$M$ = 0.172 $M_{\odot}$. Note that the mass value is very
good agreement with the mass estimate derived from the
DF00, and the effective temperature is surprisingly close
to one from the atmospheric transfer models in the
previous Section \ref{sec:TLR}. How the uncertainty in the
stellar mass affects the planet parameters will be
discussed and accounted for in Section \ref{sec:starplanet}.

\section{Star--planet parameters from direct
observables}\label{sec:starplanet}

\subsection{Using the stellar mass as the input}

Information from the transit light curves together with the
orbital solution can be combined with the stellar mass (or the
stellar radius) to fully characterize the bulk properties of
the star--planet system. These methods have been developed
over the years by different authors and we suggest reading
\citet{seager:2003}, \citet{southworth:2008} and
\citet{carter:2011} for more information on the derivations.
This subsection describes the relevant relations and the
general approach used to derive the star--planet parameters
using the stellar mass as the input (\textit{mass--input
approach}).

Assuming a circular orbit for the planet, the transit
light curve alone allows one to obtain a direct measurement
of the mean stellar density $\rho_{*, circ}$. Given a fully
Keplerian solution from the Doppler data, the stellar
density also depends on the eccentricity of the orbit and can
be written as

\begin{eqnarray}
\rho_* = \rho_{*, circ}
\left(\frac{\sqrt{1-e^2}}{1+e\sin \omega}\right)^3\,.
\label{eq:meanRho}
\end{eqnarray}

\noindent Because no prior information is required from the
orbital fit, $\rho_{*, circ}$ is a quantity typically provided
by the studies analyzing transit light curves of GJ 1214
\citep[e.g.,][]{carter:2011,kundurthy:2011}. These two studies
present the analysis of new light curves and combine them with
previous light curves from \citet{charbonneau:2009} and
\citet{sada:2010}. A weighted mean of $\rho_{*, circ}$ from
these two studies will be used in all that follows. The
detailed derivation of Eq. \ref{eq:meanRho} was first given by
\citet{seager:2003} for circular orbits, while
\citet{carter:2011} included the dependence on the
eccentricity.

With the mean stellar density from Equation \ref{eq:meanRho}
and the stellar mass from the empirical calibrations discussed
in Section \ref{sec:empmass}, one can trivially derive the
radius of the star. Using this radius and the transit depth
$R_p/R_*$ (again, a direct observable from the light curves)
one then obtains the radius of the planet. The combination of
the minimum mass (from RV) and the inclination (from transit
light curve) provides the planet's true mass, which is then
combined with the planet radius to finally derive the mean
planet density.

\begin{table}
\caption{Input parameters used to compute the Monte
Carlo distributions of the star--planet parameters}
\label{tab:input}

\begin{tabular}{lcccc}
Parameter &
Distribution &
Expected &
Standard&
Ref.\tablefootmark{A}\\
name &
type &
value &
deviation&
\\
\hline\hline\\
M$_*$ [$M_\odot$]             & Gaussian      & 0.176    & 0.009  & 1,2 \\
%R$_*$ [$R_\odot$]\tablefootmark{c} & Gaussian      & 0.200  & 0.010  & 1  \\
$\rho_{*, circ}$ [g cm$^{-3}$]& Gaussian      & 23.695   & 1.7    & 3,4 \\
a/R$_*$                       & Gaussian      & 14.62    & 0.3    & 3,4 \\
R$_p$/R$_*$                   & Gaussian      & 0.01178  & 0.001  & 3,4 \\
Inc. [deg]                    & Gaussian      & 89.19    & 0.5    & 3,4 \\
Period [days]                 & Bayesian      & 1.580400 & 1.4 10$^{-5}$ & 1 \\
K [\ms]                       & Bayesian      & 10.9     & 1.6      & 1 \\
$\lambda_0$ [deg]             & Bayesian      & 210.9    & 6.2       & 1 \\
$e\, \cos \omega$             & Bayesian      & -0.033   & 0.055    & 1 \\
$e\, \sin \omega$             & Bayesian      & -0.044   & 0.090    & 1 \\
\end{tabular}
\tablefoot{
\tablefoottext{A}{ 
(1) This work,
(2) \citet{delfosse:2000},
(3) \citet{kundurthy:2011},
(4) \citet{carter:2011}
}
}
\end{table}

To obtain realistic \textit{a posteriori} distributions,
one needs to assume realistic distributions for the
input parameters of the model. The complete list of
input parameters used at this point are given in Table
\ref{tab:input}. For all the values borrowed from the
literature, a Gaussian distribution $N\left[\mu,\sigma\right]$ 
is assumed where $\mu$ is the preferred value and 
$\sigma$ is its published uncertainty. For the
parameters derived from the orbital solution in Section
\ref{sec:rvorbit}, we directly draw samples from the MCMC
distributions generated during the orbital analysis. The
uncertainty in the stellar mass has to be accounted for at
this point. Given the precision in the distance and in the
$J, H$ and $K$ photometry, the major source of
uncertainty in the mass is due to the actual accuracy of
the DF00 calibration. This accuracy is not very well
known, but other studies suggest that it should be correct
at the 5\% level, which is the relative uncertainty we
will use for the stellar mass.

\begin{figure}[tb]
\centering
\includegraphics[width=3.2in,clip]{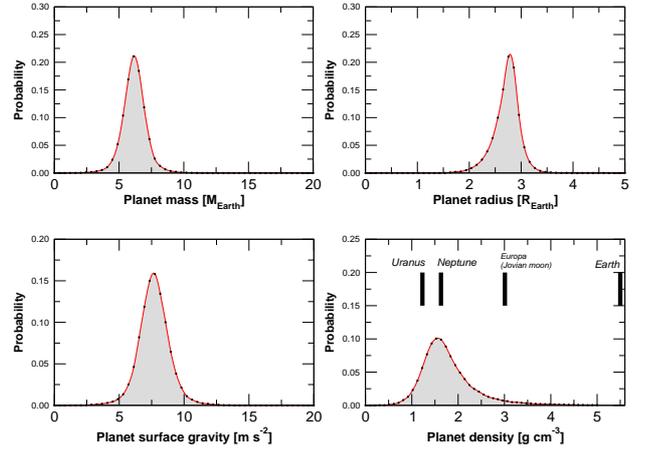}
\caption{Final probability distributions for most
significant bulk properties of GJ 1214b. The mean
densities of a few representative Solar system objects 
are marked as black vertical bars for reference. }
\label{fig:planet_values}
\end{figure}

The process of solving for all the star--planet
parameters is done for the 10$^6$ synthetic input sets. The
result is a numerical representation of the empirical
probability distributions for the derived star--planet
properties. In Figure \ref{fig:planet_values} we show
the obtained distributions concerning the planet
properties only (mass, radius, density and surface
gravity). The surface gravity of the planet can also be
derived from observables only using the prescriptions
given by \citet{southworth:2008} (combination of light
curve parameters and parameters from the orbital
solution). Table \ref{tab:starplanet}, provides the
expected values and standard deviations of the distributions
for each derived star--planet parameter.

We find that the mean density of GJ 1214b has to be
smaller than 2.40 g cm$^{-3}$ with a 95 \% confidence
level (c.l.), with 1.69 g cm$^{-3}$ being the expected value of the
distribution. An Earth-like density (e.g., $\rho>5.5$ g
cm$^3$) is thus ruled out at a $>$ 99.9 \% c.l. 
As illustrated in Fig. \ref{fig:planet_values} this
density range confirms that the planet is more similar
to a small version of Neptune, or a water dominated body
\citep[e.g., ocean planets, ]{berta:2012}, rather than a
scaled-up version of the Earth with a solid surface. This
information by itself does not solve the question about the
atmospheric composition discussed in the Introduction, but,
at least, eliminates the uncertainty in the distance of the
star in modeling of the possible planetary structure.

We want to stress that, thanks to the new distance
estimate, we now find a remarkably good agreement of the
derived stellar radius compared to the one obtained from the
SED fit in Section \ref{sec:TLR}. 

As a final test, we reproduced the star--planet parameters
using the stellar radius instead of the stellar mass. This
consists of doing the following: the stellar radius (SED fit)
combined with the mean stellar density (light curve plus RV) gives
the stellar mass. Then the stellar mass combined with the RV
observables and the inclination (light curve) provides the
planet mass. In parallel, the stellar radius with the transit
depth (light curve) gives the planet radius, which combined
with the mass finally provides the planet density. This
approach, however, has a non-obvious drawback. The mass
obtained through the stellar density formula depends as the
third power on the stellar radius ($M_*\propto
R^3$), making the stellar mass determination very sensitive to
small radius changes. Going the other way around (starting from the
mass, derive the radius), the stellar radius depends as M$^{-1/3}$, 
resulting in a much weaker dependence. For example, a $5\%$
uncertainty in the mass translates to a $1.6\%$ uncertainty in
the radius. On the other hand, a $5\%$ uncertainty in the
radius translates to a $15\%$ uncertainty in the mass. When
the uncertainty in the stellar density is also included, this
difference is even more pronounced. Therefore, we consider the
\textit{mass input} method a much more robust way of providing
a consistent picture. The updated values using the
\textit{mass input} approach for the star--planet parameters
are in Table \ref{tab:starplanet} and the marginalized
distributions for the relevant planet parameters are depicted
in Fig. \ref{fig:planet_values}

\begin{table}  % <--- column justification (center/left/right)

\caption{Suggested new star--planet parameters as derived using the stellar mass
as the only input parameter. Note that some of the distributions in
Fig.~\ref{fig:planet_values} are strongly asymetric. }

\label{tab:starplanet}
\begin{tabular}{lrrr}
Parameter &
Maximum &
Expected &
Standard\\
name &
probability\tablefootmark{\dagger} &
value &
deviation
\\
\hline\hline
Star mass [M$_\odot$]               & \textbf{0.176}
                                    \tablefootmark{a}
				            & -          & 0.0087   \\
%Star radius [R$_\odot$]             & \textbf{0.200}\tablefootmark{b} & -	 & 0.010    \\
\\
Star radius [R$_\odot$]             & 0.213 & 0.211	 & 0.011    \\
Star surface gravity [\ms]          & 1032  & 1109       & 220 \\
$\log g$ [g in c\ms]\tablefootmark{b}& 5.01  & 5.04       &  0.07    \\
Star mean density  [g cm$^{-3}$]    & 26.36 & 27.7       & 8.9      \\
\\
Planet radius [R$_\oplus$]          & 2.80 & 2.72	 & 0.24     \\
Planet mass   [M$_\oplus$]          & 6.26 & 6.19	 & 0.91     \\
Planet surface gravity [m s$^{-2}$] & 7.66 & 7.68	 & 1.19     \\
Planet mean density [g cm$^{-3}$]   & 1.56 & 1.69	 & 0.61\tablefootmark{b}     \\
\\
\multicolumn{2}{l}{Planet T$_{eq}$[K] (Bond albedo = 0)}     & 576 & 14  \\ 
\multicolumn{2}{l}{Planet T$_{eq}$[K] (Bond albedo = 0.75)}  & 407 & 11  \\
\end{tabular}
\tablefoot{
\tablefoottext{\dagger}{Maximum of the marginalized posterior distribution}
\tablefoottext{a}{Used as input}
\tablefoottext{b}{From light curve and Doppler analysis (no spectral adjustment)}
\tablefoottext{c}{$\rho_p < 2.40$ g cm$^{-3}$ with a 95\% c.l.}
\\
}
\end{table}

%\tablefoottext{b}{Derived from SED + new trigonometric parallax in Sec. \ref{sec:TLR}}

\section{Conclusions}\label{sec:conclusions}

The metallicity of GJ 1214 (and of M dwarfs in general) and
its effects on the observable fluxes (e.g., optical
fluxes) still require a better understanding, both
observationally and theoretically. Even though
spectroscopic methods in the nIR are getting better
at determining the metal content of cool stars, precise
direct measurements of their distances are still 
required for the characterization of exoplanets around
M dwarfs. At least for the stellar masses, the model
predictions from the near-infrared fluxes seem to be in good
general agreement with current empirical relations derived
from measured masses of M dwarfs in binaries (e.g.,
DF00). This is fortunate because the stellar mass is the
only input parameter required to derive the star--planet
bulk properties from direct observables. The fit of the
stellar spectral energy distributions to the
BT-Settl-2010 model grid also provides consistent
estimates in the stellar properties if infrared
photometry is used. Even with the remaining
ambiguities in the stellar parameters, both approaches
(using a stellar mass from empirical relations, or deriving
a radius from the absolute fluxes) lead to consistent
results for the star--planet bulk properties, but the
\textit{mass input} approach is preferred due to the
lower sensitivity of the method to uncertainties in
the input parameters. Several studies
have been published in recent years trying to better
characterize the GJ 1214 star--planet system. The major
source of the reported uncertainties came from directly
observable quantities that we have improved, collected,
and combined here: trigonometric parallax and
corresponding absolute fluxes, improved RV measurements,
inclusion of all of the transit observations in the derivation
of the orbital solution, additional infrared flux
measurements, and light curve observables derived from
photometric follow-up programs. 

We now find remarkable agreement of the derived star
properties obtained by comparing the $JHK$ fluxes to up-to-date
atmospheric and evolutionary models. All previous studies had
to invoke some mechanism (e.g., spot coverage) to justify the
mismatch between the predicted versus observed properties of
the star. This alone highlights the importance of obtaining
direct and accurate distance measurements of low mass stars.

At this time, the quantity that most requires further
improvement and/or independent determination is the radius of
the star. To our knowledge, this can be observationally
achieved with 3 different methods: 1) additional RV
measurements to further constrain the orbital eccentricity, 2)
direct measurement of the stellar diameter using optical/nIR
interferometry \citep[e.g.,][]{vonbraun:2012}, or 3) detection
of the secondary transit (whose instant strongly depends on
$e$ and $\omega$). We note that GJ 1214 is faint at
optical wavelengths compared to other stars observed by high
precision RV instruments \citep[e.g., HARPS or
HIRES][]{bonfils:2011,vogt:2010}. In the case of HARPS,
integrations of 45 min were required to obtain a precision of
$\sim$3.4 \ms, so a refinement of the orbital eccentricity
through additional RV measurements is time-consuming but
quite possible. On-going space--based photometric observations
in the mid-infrared (e.g., using Warm Spitzer/NASA) should be
able to detect the secondary transit soon and pin down the
orbital eccentricity to greater precision. We hope that the
updated orbital solution provided here facilitates the task of
finding such a secondary transit.

\acknowledgements We thank the referee D. S\'egransan for
useful comments that helped improving the manuscript. GA has been
partially supported by a Carnegie Postdoctoral Fellowship and by
NASA Astrobiology Institute grant NNA09DA81A. BR thanks the staff
and telescope operators of Palomar Observatory for their support.
The CAPS team (APB, AJW, GA) thanks the staff and telescope
operators of Las Campanas Observatory for the very succesful
observing runs and the Carnegie Observatories for continuous
support of the CAPS project. We thank France Allard for helpful
discussions about various topics. Part of this work is based on
data obtained from the ESO Science Archive Facility. This research
has made use of NASA's Astrophysics Data System Bibliographic
Services, the SIMBAD database, operated at CDS, Strasbourg, France.
This publication makes use of data products from the Two Micron All
Sky Survey, which is a joint project of the University of
Massachusetts and the Infrared Processing and Analysis
Center/California Institute of Technology, funded by the National
Aeronautics and Space Administration and the National Science
Foundation. This publication makes use of data products from the
Wide-field Infrared Survey Explorer, which is a joint project of
the University of California, Los Angeles, and the Jet Propulsion
Laboratory/California Institute of Technology, funded by the
National Aeronautics and Space Administration.
%\end{acknowledgements}

%\bibliography{biblio}

\clearpage

\begin{table}

\caption{Transit time observations used in the orbital fitting}
\label{tab:transittimes}

\centering                                      % used for centering table
\begin{tabular}{lcccc}
BJD     & Uncertainty & Source \\
(days)  & (days)       &        \\
\hline\hline\\
%# 1 Kundurthy 2011
%# 2 Kundurthy 2011 reanalyzed from Charbonneau et al 2009
%# 3 Carter et al. 2012
%# 4 Sada et al. 2011
2455307.892689  & 0.000263 & 1\\
2455353.724652  & 0.000311 & 1\\
2455383.752334  & 0.000264 & 1\\
2454980.748942  & 0.000417 & 1\\
2454983.909686  & 0.000401 & 1\\
2454964.944935  & 0.000788 & 2\\
2454980.748976  & 0.000264 & 2\\
2454983.909689  & 0.000228 & 2\\
2454999.713690  & 0.000253 & 2\\
2454980.748570  & 0.000150 & 3\\
2454983.909820  & 0.000160 & 3\\
2454988.650808  & 0.000049 & 3\\
2455002.874670  & 0.000190 & 3\\
2455269.962990  & 0.000160 & 3\\
2455288.928200  & 0.001100 & 3\\
2455296.830130  & 0.000230 & 3\\
2455315.794850  & 0.000230 & 3\\
2455315.794693  & 0.000080 & 3\\
2455318.955230  & 0.000170 & 3\\
2455353.723870  & 0.000180 & 3\\
2455356.884950  & 0.000150 & 3\\
2455364.787000  & 0.000150 & 3\\
2455375.849970  & 0.000130 & 3\\
2455383.752050  & 0.000130 & 3\\
2455391.654105  & 0.000059 & 3\\
2455315.793430  & 0.000420 & 4\\
2455345.821260  & 0.000140 & 4\\
2455345.821330  & 0.000370 & 4\\
2455353.723320  & 0.000360 & 4\\
2455364.786690  & 0.000290 & 4\\

\end{tabular}
\tablefoot{
(1)\citep{kundurthy:2011}
(2)\citep{kundurthy:2011} using re-analized light curves from \citet{charbonneau:2009}
(3)\citep{carter:2011}
(4)\citep{sada:2010}
}
\end{table}

\end{document}